\pdfoutput=1
\documentclass[prd,twocolumn,groupaddress,preprintnumbers,nofootinbib]{revtex4}
\usepackage{graphicx}
\usepackage{epsfig}
\usepackage{bm}
\usepackage{latexsym,amssymb,amsmath,amsfonts,amssymb,txfonts,wasysym,float}
\usepackage{scrextend}
\usepackage{mathrsfs}
\usepackage{color}

\usepackage{enumitem}

\newcommand{\postscript}[2]{\setlength{\epsfxsize}{#2\hsize}
   \centerline{\epsfbox{#1}}}

\usepackage[usenames,dvipsnames]{xcolor}
\definecolor{orange}{cmyk}{0,0.5,1,0}
\definecolor{rossoCP3}{cmyk}{0,.88,.77,.40}
\definecolor{graa}{rgb}{0.8,0.8,0.8}
\definecolor{blaa}{rgb}{0.2,0.2,0.6}

\begin{document}

\title{\color{rossoCP3} The dark dimension, the Swampland, and the origin of cosmic rays beyond the GZK barrier}

\author{Luis A. Anchordoqui}

\affiliation{Department of Physics and Astronomy,  Lehman College, City University of
  New York, NY 10468, USA
}

\affiliation{Department of Physics,
 Graduate Center, City University
  of New York,  NY 10016, USA
}

\affiliation{Department of Astrophysics,
 American Museum of Natural History, NY
 10024, USA
}

\begin{abstract}
  \vskip 2mm \noindent Very recently, it was pointed out that when basic
  ideas of the Swampland program are combined with the cosmological
  hierarchy problem (i.e. the smallness of
  the dark energy in Planck units) and together are confronted to experiment lead
  to the prediction of the
  existence of a single extra-dimension (dubbed the
dark dimension) with characteristic length-scale in the micron
range. We propose a feasible test procedure to probe the dark
dimension using the highest energy cosmic rays.
\end{abstract}
\date{May 2022}
\maketitle

\section{Introduction}

Over the past decade or so, and through many experiments, it has
become indisputable that observational data in our universe favor an effective
de-Sitter (dS) space in which the energy scale associated to the
cosmological constant $\Lambda$ is found to be $\Lambda^{1/4} = 2.31~{\rm meV}$,
corresponding to a length scale $\Lambda^{-1/4} \sim 88~\mu{\rm
  m}$~\cite{ParticleDataGroup:2020ssz}. Understanding the current period of accelerated
cosmological expansion is one of the many facets of the Swampland
program~\cite{Vafa:2005ui}. The accelerated expansion can be viewed as phenomenon that is apparently IR but
intrinsically UV, conveying a cosmological hierarchy induced by the
smallness of
  the dark energy in Planck units: $\Lambda \sim 10^{-122} M_{\rm
    Pl}^4$.

  It was recently pointed out that combining the Swampland program with the
  cosmological hierarchy problem and confronting these basic ideas to
  experiment lead to the prediction of the existence of a single
  extra-dimension (dubbed the dark dimension), with characteristic
  length-scale in the micron range~\cite{Montero:2022prj}. The dark
  dimension carries with it the emergence of the so-called ``species
  scale'', where gravity becomes strongly coupled~\cite{Dvali:2007hz,Dvali:2007wp}.

It was also pointed out in~\cite{Montero:2022prj}
that the species scale is tantalizingly close to the
energy above which the Telescope Array (TA) and the Pierre Auger
collaborations found conclusive evidence for a sharp cutoff of the flux of ultra-high-energy cosmic rays (UHECRs)~\cite{HiRes:2007lra,PierreAuger:2008rol}. In this paper
we elaborate on this happenstance and show how future UHECR measurements
can provide a direct test of the dark dimension.

The paper is organized as follows. In Sec.~\ref{sec:2} we review
general aspects of the effective low energy theory that is consistent
with current observations of dark energy while naturally satisfying
conjectured swampland constraints. In the spirit of~\cite{Montero:2022prj}, in
Sec.~\ref{sec:3} we argue that the dark dimension limits the maximum energy to which UHECRs can
be accelerated and that this theoretical limit is of the same order of magnitude that the upper limit for energy at which cosmic rays have experimentally been detected.
After that we propose a possible test procedure to distinguish among
theoretical models that successfully describe the sharp cutoff
observed in the UHECR spectrum, including the one 
associated to the dark dimension. In Sec.~\ref{sec:4} we draw the final
conclusions and provide an additional argument to discriminate among
theoretical models describing the high-energy end of the cosmic ray spectrum.

\section{UV constraints on IR physics}
\label{sec:2}

The Swampland program focuses on discriminating effective field
theories (EFTs) which can be completed into quantum gravity in the UV
from those which cannot~\cite{Vafa:2005ui}. In theory space, the
border circumventing EFTs that are compatible with string theory from the ``swampland''
is characterized by a set of conjectures on the properties that these
EFTs should have/avoid in order to allow a consistent completion into
quantum gravity. There are many swampland conjectures in the market,
indeed too many to be itemized here, and so we refer the reader to comprehensive reviews~\cite{Palti:2019pca,vanBeest:2021lhn}.

Word for word and letter for letter, the {\it distance swampland
  conjecture} suggests that an infinite tower of modes becomes
exponentially light when approaching a point that is at infinite
proper distance in field space~\cite{Ooguri:2006in}. Expressed in the
form of the {\it anti-dS (AdS) distance
  conjecture}~\cite{Lust:2019zwm}, it reveals that the mass of the
tower of states is related to the magnitude of the cosmological
constant. To be precise, the mass scale $m$ behaves as
$m \sim |\Lambda|^\alpha$, as the negative AdS vacuum energy
$\Lambda \to 0$, with $\alpha$ a positive constant of ${\cal O} (1)$.
In~\cite{Lust:2019zwm} also some implications of the AdS distance
conjecture for de Sitter space were discussed, namely, when assuming
this scaling behavior to hold in dS (or quasi dS) space with a
positive cosmological constant, approaching $\Lambda = 0$ will also
lead to an unbounded number of massless modes. Now, these
considerations when merged with the cosmological hierarchy
problem spontaneously point to a particular corner of the string
landscape, associated to an asymptotic region of the field space.

Now, consistency of large-distance black hole physics in the
presence of $N$ elementary particle species imposes a bound on the
gravitational cutoff of the EFT, and the fundamental length is no
longer $l_{\rm Pl} = M_{\rm Pl}^{-1}$, but rather~\cite{Dvali:2007hz,Dvali:2007wp}
\begin{equation}
l_N = \sqrt{N} \, M_{\rm Pl}^{-1} \, .
\label{lN}
\end{equation}
The origin of (\ref{lN}) can be traced back using different
arguments, we follow here the reasoning given in~\cite{Antoniadis:2014bpa} based on quantum
information storage~\cite{Dvali:2008ec}. Consider a pixel of
size $L$ containing $N$ species storing information. The energy required to localize $N$ wave functions
is found to be $N/L$. This energy can be associated to a Schwarzschild radius
$r_s = N/ (LM_{\rm Pl}^2)$, which must be smaller
than the pixel size if we want to avoid the system to collapse into a
black
hole. Now, $r_s \leq L$, implies there is a minimum size $l_N \equiv L_{\rm min} = \sqrt{N} \, M_{\rm
  Pl}^{-1}$ associated to the EFT UV cutoff (a.k.a. species scale).  The particle species we want to consider here are the Kaluza-Klein
(KK) excitations of the graviton (and other possible bulk modes) given
by $N \sim R_\perp^{n} l_N^{-n}$,
up to energies of order
\begin{equation}
 M_{\rm UV}  =  m^{n/(n+2)} M_{\rm Pl}^{2/(n+2)},
\end{equation}
where $n$ is the number of decompactifying  dimensions of radius $R_\perp \sim m^{-1}$ and $M_{\rm UV} =
l_N^{-1}$ is the species scale that corresponds to
 the Planck scale of the higher dimensional theory.

 The tower of KK gravitons leads to significant deviations from
 Newton's gravitational inverse-square law at the energy scale
 $m$. Thus far no such deviations have been found in the
 short length-scale ($\sim 30~\mu {\rm m}$) regime~\cite{Lee:2020zjt},
 and so we arrive at the bound $m \agt 6.6~{\rm meV}$. Requiring this 
 experimental bound to be consistent with the theoretical bound from
 the swampland conjectures suggests $\alpha =1/4$, and so the mass
 scale of the KK modes in the tower is estimated to be
 $m \sim \Lambda^{1/4}/\lambda$, with
 $10^{-4} \alt \lambda \alt
 10^{-1}$~\cite{Montero:2022prj}. Consistency with neutron star
 heating by the surrounding cloud of trapped KK
 gravitons~\cite{Hannestad:2003yd} requires $n=1$~\cite{Montero:2022prj}. In closing, we note that for $\lambda \sim
10^{-3}$ we have  $R_\perp  \sim 1~\mu{\rm m}$ and physics becomes
strongly coupled to gravity at $M_{\rm UV} \sim \lambda^{-1/3} \Lambda^{1/12} M_{\rm Pl}^{2/3} \sim
10^{10}~{\rm GeV}$.

\section{Evidence of UV physics in Auger data?}
\label{sec:3}

The TA and the Pierre Auger collaborations found conclusive evidence that
the cosmic ray flux drops precipitously for energies $E \agt 10^{10.6}~{\rm
  GeV}$~\cite{HiRes:2007lra,PierreAuger:2008rol}.  However, such an
observation is still not conclusive on the origin of the
suppression. There are two competing models to explain the
observed suppression: {\it (i)}~the Greisen-Zatsepin-Kuz’min (GZK) effect due to UHECR interactions with the cosmic
microwave background~\cite{Greisen:1966jv,Zatsepin:1966jv} and {\it (ii)}~the limiting acceleration energy
hypothesis~\cite{Allard:2008gj}, wherein it is postulated that the
``end-of steam'' for cosmic accelerators is coincidentally near the
putative GZK cutoff, with the exact energy cutoff determined by
data. In the spirit of~\cite{Montero:2022prj}, herein we explore a third elegant 
explanation, in which the observed suppression is controlled by the
species scale of the 5D theory.

In analogy with the electromagnetic difraction radiation emitted by a
charge near a metallic grating (a.k.a. Smith-Purcell
effect~\cite{Smith:1953sq}) we would expect UHECRs to emit
gravitational difraction radiation (GDR) due to the lumpiness of spacetime
at small scales~\cite{Cardoso:2006nz}. More concretely, GDR on the visible (Standard Model) brane could be
generated by inhomogeneities on a second hidden brane due to
bulk-brane interactions and brane fluctuations~\cite{Bando:1999di}. As
an illustration of this gravitational phenomenon, we compute the power
radiated by UHECRs in the presence of a hidden brane, with typical
longitudinal perturbations of length scale $\delta_\parallel$ and
transverse perturbations of length scale $\delta_\perp$. These
perturbations are modeled with a $\delta_\parallel$-periodic lamellar
grating with rulings of width $\epsilon$ perpendicular to the particle
direction of motion.\footnote{The particulars of the Smith-Purcell
  effect for a single grating with these characteristics have been analyzed in~\cite{Shibata}.} The GDR energy loss per unit distance is found to be
\begin{equation}
\frac{d \ln E}{dx} \sim - 8 \pi \ E \ R_\perp \ l_{\rm Pl}^2 \
\frac{\delta_\perp^2}{\delta_\parallel^5} \ e^{-2\pi R_\perp/(\Gamma \delta_\parallel)} \  \left[1 + {\cal O} (\delta_\perp/\delta_\parallel) \right] \,,
\label{GDR}
\end{equation}
where $\Gamma$ is the UHECR Lorentz boost, and where we have set
$\delta_\parallel = 2\epsilon$ and of course assumed that
$\delta_\perp/\delta_\parallel \ll 1$~\cite{Cardoso:2006nz}.

For fiducial values $\delta_\perp \sim 0.1 \delta_\parallel$ and
$\delta_\parallel \sim  10^{-3} R_\perp$, the exponential in (\ref{GDR}) can
be set to unity for $\Gamma \agt 10^{5}$ and therefore the GDR energy loss
rate is the same for all baryonic species. (\ref{GDR}) can be recast as
\begin{eqnarray}
 \frac{d\ln E}{dx} & \sim & -8\pi \frac{E}{M_{\rm UV}^3} \
  \frac{\delta_\perp^2}{\delta_\parallel^5} \nonumber \\
& \sim & - 0.1 \ \left(\frac{E}{10^{10}~{\rm GeV}}\right)~{\rm
      Mpc}^{-1} \, .
\end{eqnarray}
For $E \sim 10^{11}~{\rm GeV}$, the GDR fractional energy loss rate,
${\cal O} ({\rm Mpc}^{-1})$, is more severe than the GZK energy
losses, and comparable to the average distance between Galaxies. The GDR can then be interpreted as limiting the maximum
energy to which UHECRs can be accelerated, and the source spectra can
be effectively modelled as~$\propto E^{-\gamma} \ \exp \{-E/M_{\rm UV}\}$, where $\gamma$ is a
free parameter.

Because both the GZK
    prediction and the GDR and limiting-acceleration-energy  
hypotheses accommodate the same rate in the mean, it is
difficult to discriminate between them. Nevertheless, as we show in
what follows, a discrimination becomes feasible by analyzing UHECRs
beyond the onset of the suppression.

The Pierre Auger Collaboration reported a $4\sigma$
significance for a correlation between the arrival direction of cosmic
rays with $E \agt 10^{10.6}~{\rm
  GeV}$ and a model based on a catalog of bright starburst
galaxies~\cite{PierreAuger:2018qvk, PierreAuger:2021rfz}. When TA data are included in the analysis the
correlation with starburst galaxies is mildly stronger than the
Auger-only result, with a post-trial significance of
$4.2\sigma$~\cite{TelescopeArray:2021gxg}.

Because starburst galaxies
are mostly nearby, the UHECR flux attenuation factor due to GZK
interactions {\it en route} to Earth turns out to be negligible. Indeed, taking
into account attenuation about 90\% of the accumulated flux from
starburst model emerges from a 10~Mpc-radius region. This implies that
for each source, the shape of the observed spectrum should roughly match
the emission spectrum by the starburst. Constraints based on the
isotropic gamma-ray background at TeV measured by {\it Fermi} LAT~\cite{Fermi-LAT:2015otn} seem to support the association of UHECRs and nearby starburst galaxies~\cite{Liu:2016brs}.

On the one hand,  if the flux suppression were primarily caused by the limiting
acceleration energy at the sources rather than by the GZK effect, in
the absence of phenomena engendered by the dark dimension, we would
expect a high variance of parameters characterizing the UHECR source
spectra $\propto E^{-\gamma} \ \exp \{-E/E_{p,_{\rm
      max}}\}$, reflecting the many different properties inherent to
  the acceleration
  environments in the market~\cite{Anchordoqui:2018qom}. A point worth noting at
  this juncture is that the commonly adopted
  source spectra  assume the validity
  of the Peters cycle: $E_{\rm max} (Z) = Z \, E_{p,{\rm max}}$, where $E_{p,_{\rm
      max}}$ is the 
maximum energy to which protons can be accelerated and $Z$ is the
charge of the UHECR in units of the proton charge~\cite{Peters:1961}.  On the other hand,
as we discussed above, the cutoff spectra must
be universal if physics becomes strongly coupled to gravity above
about $10^{10}~{\rm GeV}$. This difference could be a useful tool to
discriminate among the models.

\begin{figure}[tbp]
    \postscript{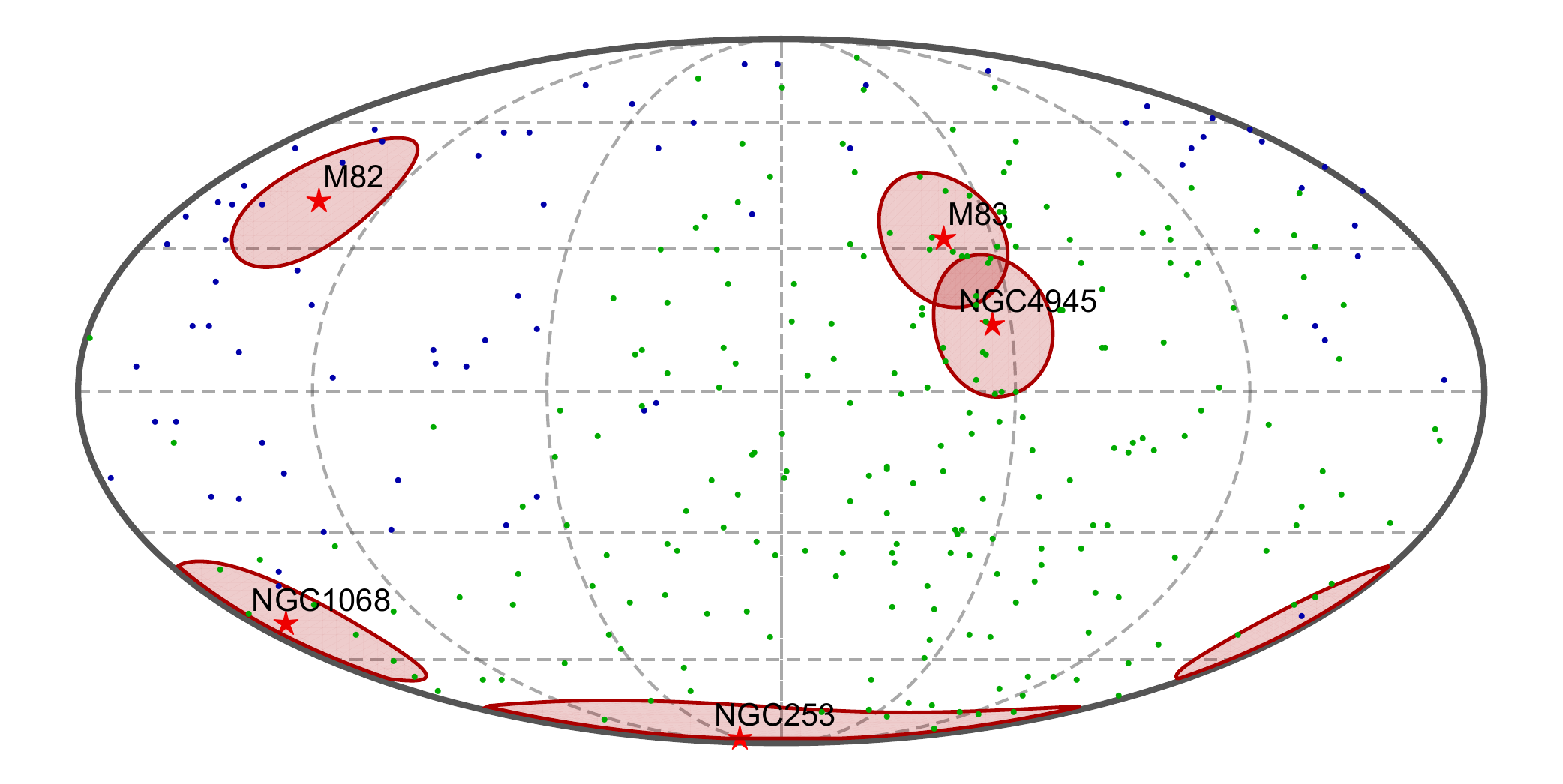}{0.9}
\caption{Skymap in Galactic coordinates of the arrival directions of
  231 cosmic rays with $E > 10^{10.72}~{\rm
    GeV}$ (green dots) and the 72 cosmic rays with $E >
  10^{10.76}~{\rm GeV}$ (blue dots); see text for details. The locations of the four starbursts yielding the
  dominant signal of the observed anisotropy are indicated by stars
  and the associated shaded regions
  delimit $15^\circ$
 circular windows around the starbursts.}
\label{figura}
\end{figure}

\begin{table}
  \caption{Results of the likelihood fit.}
\begin{tabular}{cccccc}
\hline
\hline
Starburst   & Experiment & Events &    $\gamma$  &  $\gamma_{\rm min}$
  &    $\gamma_{\rm max}$ \\
\hline
NGC 4945	   & Auger	&	14	&	6.8	&	5.4	&	8.5\\
M83		 &  Auger	&	13	&	4.6	&	3.7	&	5.7\\
~~~~NGC 253~~~~ & ~~~~Auger~~~~	&	~~~~8~~~~	&	~~~~4.8~~~~	&	~~~~3.6~~~~	&	~~~~6.4~~~~\\
NGC 1068	 & Auger	&	8	&	4.9	&	3.7	&	6.4\\
NGC 1068	 & TA	&	2	&	3.9	&	2.3   &	6.5\\
M82		& TA	&	3	&	5.3	&     3.3    &	8.3 \\
\hline
\hline
\end{tabular}
\label{tabla}
\end{table}   

Next, we demonstrate one possible test procedure using public data. In
our analysis we
consider 231 cosmic rays with $E  > 10^{10.72}~{\rm GeV}$ and incident zenith angle $\theta < 80^\circ$ detected by the Pierre
Auger Observatory during 1/1/2014 and
3/31/2014~\cite{PierreAuger:2014yba} as well as 72 cosmic rays with
$E > 10^{10.76}~{\rm GeV}$ and $\theta < 55^\circ$ detected by
TA during 05/11/2008 and 05/04/2013~\cite{TelescopeArray:2014tsd}. For
the four starburst galaxies dominating the anisotropy signal, we define an circular window around their location in  the sky with
angular radius of $15^\circ$, as shown in Fig.~\ref{figura}. Such an 
angular radius serves just as an orientation to illustrate the test
procedure. Because $E > M_{\rm UV}$ for all of the events in the sample,
we simply assume unbroken power law spectra $\propto
E^{-\gamma}$ to describe the sharp fall off and we carry out a maximum likelihood estimation of the 
spectral index at the sources. Our results are encapsulated in
Table~\ref{tabla}, where we give the number of events within each source
circle together with the values of $\gamma$ maximizing the likelihood
and 
the $68\%$ confidence level intervals $[\gamma_{\rm min},\gamma_{\rm
  max}]$. The starburst 
individual spectra are very steep, reflecting the suppression of the
UHECR flux, and are all consistent with $\gamma = 5$ within $1\sigma$,
supporting universality. We conclude that the analyzed database does not
constrain $M_{\rm UV}$.

Continuing operation of Auger should yield a significance level of $5\sigma$ of the starburst hypothesis by the end
of 2025 ($\pm 2$ calendar years)~\cite{Coleman:2022abf}. Such a high-statistics database
could be much more impactful in testing  whether the observed cutoff
spectra of the starbursts are universal in origin.

\section{Conclusions}
\label{sec:4}

Motivated by principles from the Swampland program it was recently
proposed that our Universe may be close to an infinite distance
limit~\cite{Vafa:2005ui}. We have investigated some phenomenological
consequences of this intriguing proposal that successfully addresses the cosmological hierarchy
problem. We have conjectured that it can provide an elegant explanation of the 
the abrupt cutoff observed in the UHECR flux. We have shown that this explanation is testable with present and near-future UHECR observatories. 

We end with an observation: cosmogenic neutrinos~\cite{Berezinsky:1969erk} could provide a
smoking-gun signal of the GZK effect, but no neutrino has ever been observed above $10^7~{\rm GeV}$.

\section*{Acknowledgments}

The research of LAA is supported by
the U.S. NSF grant PHY-2112527 and NASA grant 80NSSC18K0464.

\end{document}